\documentclass[%
 reprint,
 showkeys,
nofootinbib,
 amsmath,amssymb,
 aps,
 pra,
]{revtex4-2}

\usepackage{graphicx}
\usepackage{dcolumn}
\usepackage{bm}
\usepackage[hidelinks]{hyperref}
\usepackage{nicematrix} 
\usepackage{booktabs}

\begin{document}


\title{A mark and recapture perspective on vaccination touchpoints}
\author{Niket Thakkar}
\email{niket.thakkar@gatesfoundation.org}
\homepage[\\]{https://github.com/NThakkar-IDM/recapture}
\affiliation{%
 The Gates Foundation's Institute for Disease Modeling\\
 Seattle, Washington 98109
}%
\date{\today}

\begin{abstract}
This paper considers large-scale vaccination campaigns, a major platform for vaccine access in a lot of the world, as a recapture estimate of the target population marked by routine immunization. Framing the campaign as a measurement, we learn about its properties, including the campaign's coverage of the target population and some implied sampling properties of post-campaign coverage surveys (PCCSs), the current gold-standard in implementation quality measurement.

We develop this idea in the context of the 2023 measles campaign in Kano State, Nigeria, where we have detailed implementation data collected by vaccination teams involved in that effort. Looking specifically at the teams' tally sheets, the daily records of who they vaccinated, we find significant discrepancies between the recapture estimates and those from the corresponding PCCS. Exploring a variety of bias models applied to both the tally sheets and the PCCS helps clarify how anecdotal issues from the field relate to this discrepancy. 

Overall, we find that the tally sheets, despite being an unorthodox population sample, provide a tractable perspective on implementation and measurement, one that's in principle available nearly instantly and at high resolution for any vaccination touchpoint. 
\end{abstract}
\maketitle


\section{Introduction}

It's generally recognized that a commitment to improving global health is also a commitment to measuring the properties of interventions \cite{lindstrand2024implementing}. Unfortunately, the consensus tends to disintegrate from there. What needs to be measured, to what precision, how to do it, who it's for (and so who should pay for it) are all open questions in a broadly chaotic measurement environment. 

Measles vaccine delivery is a concrete example. In the highest-burden settings, routine vaccination services are inaccessible or unreliable, leaving gaps in immunity that have to be closed by dedicated vaccination campaigns every few years \cite{world2016planning}. In those efforts, medical teams bring vaccines to underserved areas and try to immunize kids within a predetermined target population. Designing these efforts to better reach susceptible individuals before the virus does is a major global health initiative \cite{mri_strat}, one that requires high-quality measures of intervention impact, comprehensiveness, and failure-modes.

The accepted standard for measuring campaign quality is the post-campaign coverage survey \cite{world2016planning,DanovaroHolliday2024}. PCCS is a 2-stage cluster survey where individuals in the target population are asked about their campaign experience, including if they received a vaccine at all. While useful in principle, in practice PCCS is underpowered at the granular scale relevant to implementation, it's rarely conducted in a reasonable time frame, and often its findings are colloquially reduced to just a coverage number and then forgotten. It's becoming clear that PCCS isn't satisfying programmatic measurement needs. In fact, most countries that conducted a campaign since 2020 haven't even bothered with a PCCS \cite{DanovaroHolliday2024}. 

For a complex intervention like a vaccination campaign it seems unwise to rely so heavily on one measurement platform. Ideally, measurement methods offer light touch, timely, accurate, and interpretable insights into intervention details, which is a lot to ask of a single approach. As campaigns progress with no measurement plans at all, the toolbox is starting to feel pretty empty.

Along those lines, the main idea in this paper is to make better use of the implementation records already collected in a vaccination campaign. We focus specifically on the tally sheets \cite{higgins2025recording}, Fig. \ref{fig:tally} is a blank example, literal paper records of who vaccination teams encounter as they work. In exploring the data photographed and digitized during the 2023 measles campaign in Kano State, Nigeria \cite{grid3_pilot}, we show that campaign coverage can be estimated directly from the tally sheets in a principled Bayesian inference scheme, making use of prior information on routine immunization coverage and target population size. The approach is analogous to mark and recapture methods common in ecology \cite{Chapman1954,chao2001applications}, and so we call this the recapture coverage estimate.

In the specific example considered, the recapture coverage estimate is in disagreement with the corresponding PCCS estimate. Two quiet but common narratives in global health are that tally sheets are at least partially fabricated to increase campaign scale and that PCCS sampling is biased to increase coverage in response to political pressures on performance. We consider these hypotheses quantitatively in our example, and we show that an unrealistically structured bias in tallying would be necessary to corroborate the PCCS.

This paper does not fully settle the issue of campaign coverage and target population size in 2023 in Kano. But it does demonstrate that implementation records offer timely and interpretable insight into campaign coverage. As our understanding develops, that could free the PCCS to focus on other implementation details, and we conclude with some general thoughts to that effect.

\section{Four confusing numbers}

The 2023 measles vaccination campaign in Kano State took place over 9 days in mid December. In the planning phase, the Gates Foundation funded GRID3 to support implementation in a variety of ways, including a pilot study where paper records were photographed and hand-digitized by GRID3 employees in 6 local government areas (LGAs, the Nigerian administrative level 2 unit, essentially districts) considered representative of the state\footnote{It's not easy to verify this claim in this context. The LGAs in question cover $15.6\%$ of the state's population. Half of the LGAs are urban, the other half are rural in the sense that they're outside Kano City. Recent LGA-level estimates \cite{haeuser2025global} of RI coverage have the 6 LGA average deviating from the state average by $<1\%$. In any case, since the focus of this paper is the method, not any specific campaign coverage estimate, we don't explore potential ecological fallacies in much detail.}. The tally sheet photographs in particular were taken on a nightly basis in a direct interaction with vaccination team supervisors \cite{grid3_pilot}. 

\begin{figure}[b]
\centering\includegraphics[width=\linewidth,trim={2cm 0 2cm 0},clip=True]{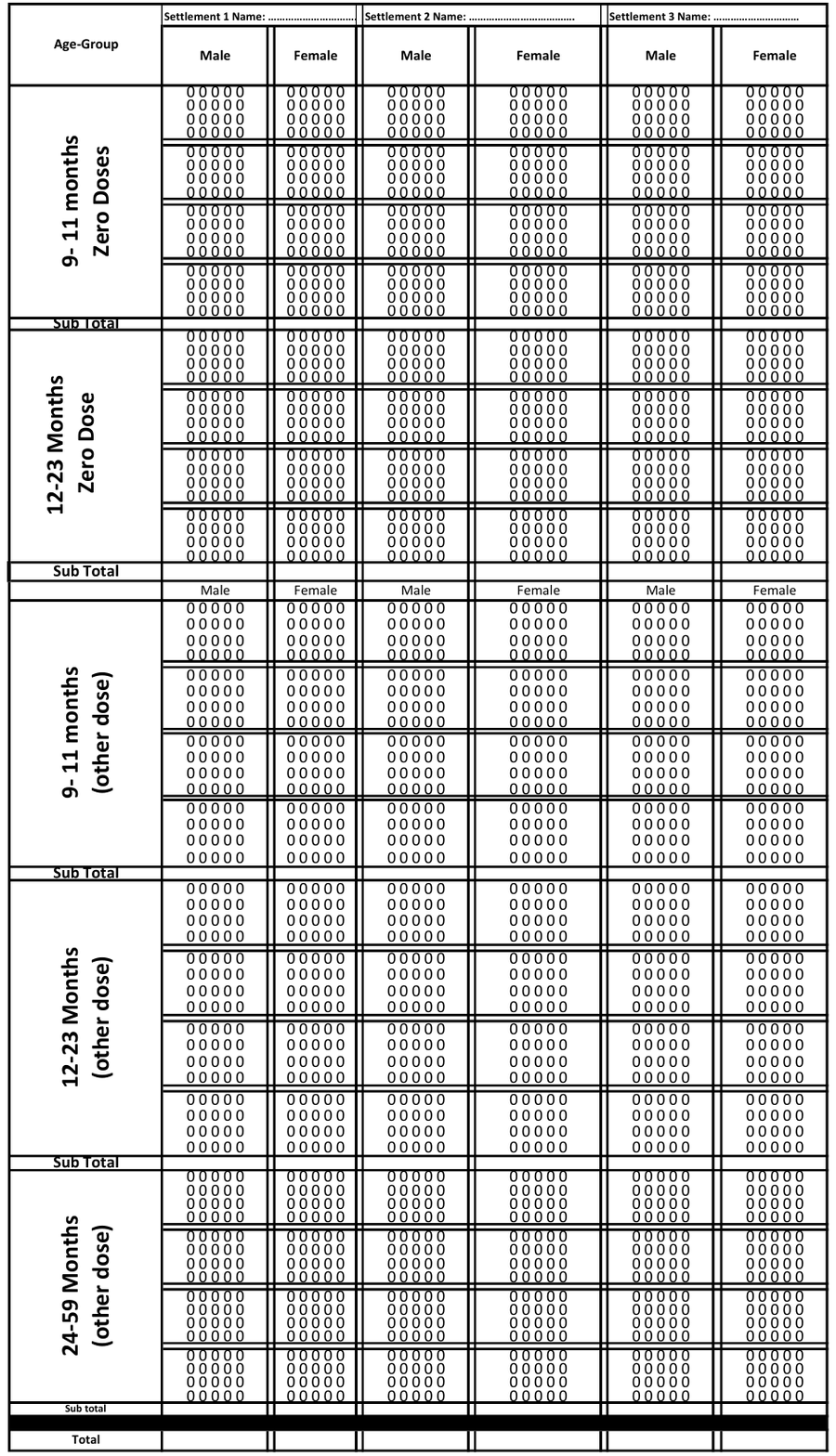}
\caption{\label{fig:tally} A blank tally sheet from Nigeria circa 2023. Vaccinators tick off the relevant bubbles as they work, compiling a histogram each campaign day.}
\end{figure}

For our purposes, the implementation data is essentially a collection of counts by vaccination team every campaign day, taken from records like that in Fig. \ref{fig:tally}. While the tallies are disaggregated by sex and age, we'll focus specifically on the 12 to 23 month-olds across their caregiver reported status of having had measles vaccine before (``other dose'', OD) or not (``zero dose'', ZD). In Kano, across the 6 LGAs, there were $204137$ 1 year-old tallies in total, with $189321$ reporting OD. Since the previous campaign in Kano was in 2021, these caregivers must have been referring to vaccines delivered through the routine immunization (RI) system.

With that thought in mind, we might take tallied 1 year-olds as a sample of RI coverage and conclude that $92.7\%$ of kids get measles vaccine through RI in Kano. In the same year though, the Demographics and Health Survey (DHS) found that RI coverage among 1 year-olds was $38.6\%$ \cite{nigeria2023dhs}. Meanwhile, for this campaign, the PCCS estimated that $89\%$ of the target population was covered, with $11\%$ of them getting their first ever measles vaccine from the intervention \cite{nigeria2023pccs}. These numbers are collected in Table \ref{tab:data} with some sample size and design details.

\begin{table}
\caption{%
Routine, campaign, and post-campaign measles vaccine coverage data from Kano State Nigeria, circa 2023
}\label{tab:data}
\begin{ruledtabular}
\begin{tabular}{lcc}
\textrm{Source}&
\textrm{Coverage}&
{Sample size}\\
\colrule
DHS\footnote{The 2023 DHS does not yet have a published design effect. We take $D_{e}=1.303$ from the 2018 DHS throughout \cite{nigeria2018dhs}.}  (12--23mo.) & .386 & 531\\
Tally sheets (12--23mo.) & .927 & 204137\\
PCCS\footnote{\label{pccs_note}Approximately 107 of the 600 samples are 1-year-olds, and $D_{e} = 8$ for overall coverage \cite{nigeria2023pccs}. We take estimates across ages as representative of 1 year-olds but with higher uncertainty.} (9mo.--5yr.) & .89 & 600\\
PCCS with ZD at the campaign\footnote{$D_{e} = 2$. Estimates across all states suggest this could be $\approx.15$ for 1 year-olds, higher, but still within the Kano interval \cite{nigeria2023pccs}.} & .11 & 600
\end{tabular}
\end{ruledtabular}
\end{table}

It's not clear how the pieces fit together. For example, taking the PCCS and tally sheets at face value and assuming every missed child is ZD, we'd estimate RI coverage is $82.5\%$ at a minimum, more than $22$ standard deviations from the DHS. Meanwhile, accepted population estimates like that from GRID3 \cite{grid3website} put approximately $144$k 1 year-olds in the 6 LGAs, so we sometimes see reports that campaign coverage is an impressive $\approx 140\%$. 

The usual thing to do at this stage is to argue that the DHS and PCCS have rigorous sample designs while the tally sheets do not. Something must be wrong with the latter, and they should be ignored. As we'll see, there is probably something wrong with the tally sheets. But they shouldn't be ignored.

\section{Mark and recapture}

\begin{figure*}
\centering\includegraphics[width=\linewidth]{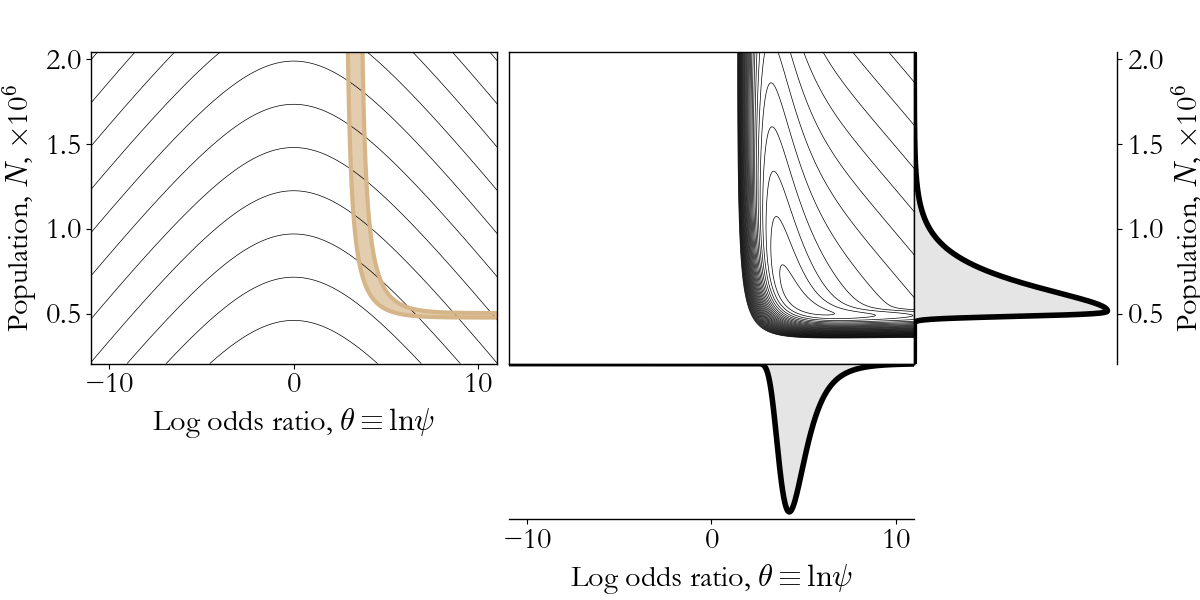}
\caption{\label{fig:inf1} The tally sheets at face value. (Left) The joint prior distribution $p(N,\theta)$ (log contours in black) gives the $N,\theta$-space some basic structure. The likelihood (tan) highlights the ridge consistent with the data. (Right) Their combined contributions are evident in $p(N,\theta\,|\,\Delta)$ (log contours at the same levels), which can be used to compute the relevant marginals (grey).}
\end{figure*}

The size of the target population, that is the number of 1 year-olds in the 6 LGAs, is a fundamental uncertainty underlying the discordance above. Mark and recapture methods are an established  approach to estimating population sizes based on relationships between distinct samples. They inspire a line of thinking for us in Kano.

For the unfamiliar, the canonical \cite{Chapman1954} mark-recapture example is something like this: An ecologist wants to know how many fish are in a lake. She first goes to the lake, catches $10$ fish, and marks them in some durable way. Then she puts them back into the water. When she returns the next day, she catches $20$ fish, of which $4$ are marked. Since the expected number of marked fish in $20$ samples from a large population of $N$ fish is $20\times(10/N)$, she might estimate that there are $N = 50$ fish in the lake, the so-called Lincoln-Peterson \cite{lincoln1930calculating,petersen1896yearly} estimate.

Our situation is a variant where we have some uncertainty in the number of individuals marked. Specifically, in Kano, the DHS tells us that $38.6\%$ of the population are ``marked'' by vaccine through RI, and in the campaign, teams found a sample of $204$k children, of which $92.7\%$ were marked. If we assume that the campaign exhausted the population covered by RI, that would imply there are $204\text{k}\times(92.7/38.6)\approx490$k 1 year-olds in the population, and campaign coverage was about $42\%$. 

This quick calculation is the basic idea in what follows. It turns out, with some care, we can leverage prior information to avoid assumptions about campaign coverage in OD children. Moreover, in framing the problem probabilistically, we can quantify uncertainty more thoughtfully throughout, giving us some perspective on the PCCS estimates in Table \ref{tab:data} as well.

To get started, consider a population of $N$ 1 year-olds. Conditional on $N$, we can define a length $N$ binary vector $\mathbf{R} = (R_1, R_2, ..., R_N)$ representing each 1 year-old's RI status just before the campaign, with $R_i = 1$ if the $i$th child received one or more measles vaccines through RI. The population has some true RI coverage
\begin{equation}
    C_N(\mathbf{R}) = \frac{1}{N}\sum_{i=1}^N R_i,\label{eq:true_cov}
\end{equation}
a population average based in this case on $N$ samples, that is, a census for a given $N$.

For our purposes, the DHS is an estimate of Eq. \ref{eq:true_cov}, and we can retrieve the corresponding average and variance directly from the survey report. But for completeness and clarity, we treat the DHS as another length $N$ binary vector $\mathbf{S}^0 = (S_1^0, S_2^0, ..., S_N^0)$, where $S^0_i = 1$ means a child was sampled in the DHS, and $\sum_i S^0_i = n_0$ is a known constraint called the sample size. The relevant estimate in our case is
\begin{equation*}
    C_{n_0}(\mathbf{R}) = \frac{1}{n_0} \sum_{i=1}^N S_i^0 W_i^0 R_i,
\end{equation*}
a weighted average where $W_i^0$ is the $i$th child's normalized survey weight. The DHS sample is constructed for these averages to be unbiased and $\text{E}[C_{n_0}(\mathbf{R})] = C_N(\mathbf{R})$, where the expected value is taken over $p(\mathbf{S}^0\,|\,n_0)$, a joint distribution sometimes called the sample design. Meanwhile, for binary $\mathbf{R}$, the variance $\text{V}[C_{n_0}(\mathbf{R})] \approx (D_e^0/n_0)\text{E}[C_{n_0}(\mathbf{R})](1-\text{E}[C_{n_0}(\mathbf{R})])$, where $D_e^0$ is called the design effect, quantifying variance inflation relative to a less economical, large $N$ simple random sample \cite{cochran1977sampling}.

Along similar lines, the campaign is a distinct sample, again a length $N$ binary vector $\mathbf{S}^1 = (S_1^1, S_2^1, ..., S_N^1)$ where $S_i^1 = 1$ means the $i$th child received a campaign vaccine and $\sum_i S^1_i = n_1$ is known from the tally sheets. The sample is unorthodox in the sense that it has unknown design. In other words, we need a reasonable and tractable model for statistics of $p(\mathbf{S}^1\,|\,n_1)$. 

To make progress, we consider $R_i$ as a correlate of the $i$th child's campaign participation, and we can hypothesis test against a sampling model with two distinct sub-populations, those with and without routine measles vaccine. Specifically, let
\begin{align*}
    k_1 \equiv \sum_{i=1}^N S^1_iR_i &\sim \text{Binom}\left\{NC_N(\mathbf{R}),\pi_1\right\},\\
    k_0 \equiv \sum_{i=1}^N S^1_i(1-R_i) &\sim \text{Binom}\left\{N(1-C_N(\mathbf{R})),\pi_0\right\},
\end{align*}
such that $k_0+k_1 = n_1$ and both $\pi_0$ and $\pi_1$ are unknown sampling probabilities. Standard manipulations give the more concise form
\begin{equation}
    p(k_1|n_1,\mathbf{R}) \propto \binom{NC_N(\mathbf{R})}{k_1}\binom{N(1-C_N(\mathbf{R}))}{n_1 - k_1}\psi^{k_1},\label{eq:nch_dist}
\end{equation}
where $\psi \equiv \pi_1(1-\pi_0)/\pi_0(1-\pi_1)$ is the odds ratio of being sampled based on RI status, and conditionality on $N$ is implicit in the conditionality on $\mathbf{R}$. For general $\psi \neq 1$, this construction is sometimes called Fisher's noncentral hypergeometric distribution \cite{EisingaPelzer2011}. Its properties and large $n_1$ asymptotic approximations to its mean and variance are discussed in Appendix \ref{ap:A}.

In the context of Eq. \ref{eq:nch_dist}, the tally sheet RI coverage estimate is $C_{n_1}(\mathbf{R}) = k_1/n_1$. As a result, the observed difference, $\Delta \equiv C_{n_0}(\mathbf{R}) - C_{n_1}(\mathbf{R})$ has statistics
\begin{align*}
    \text{E}[\Delta] &= \text{E}[C_{n_0}(\mathbf{R})]-\frac{1}{n_1}\text{E}[k_1],\\
    \text{V}[\Delta] &= \text{V}[C_{n_0}(\mathbf{R})]+\frac{1}{n_1^2}\text{V}[k_1],
\end{align*}
where in the second line we assume the two samples are independent. From Appendix \ref{ap:A}, $\text{V}[k_1] \sim \mathcal{O}(n_1)$, so the DHS variance dominates when $n_1 >> n_0$, and $p(\Delta\,|\,N,\psi)$ is Gaussian to a good approximation when we're thinking about a scale where the DHS is well-powered.

Under that assumption, Bayes' theorem gives us an approachable, 2-dimensional inference problem characterizing the campaign,
\begin{align}
    \begin{split}
    p(N,\theta\,|\,\Delta) &= \frac{p(\Delta\,|\,N,\theta)p(N,\theta)}{p(\Delta)}\\
    &\approx \frac{\mathcal{N}(\Delta\,|\,\text{E}[\Delta],\text{V}[\Delta])p(N)p(\theta)}{p(\Delta)},
    \end{split}\label{eq:inf}
\end{align}
where $\mathcal{N}$ is a Gaussian distribution, and we've chosen to work in terms of $\theta \equiv \ln\psi$ to facilitate the numerical integration needed to estimate $p(\Delta)$ and solve the inference problem. Note that all the distributions in Eq. \ref{eq:inf} are implicitly conditional on the sample sizes $n_0$ and $n_1$. 

In Appendix \ref{ap:B}, we show that the maximum entropy principle \cite{Shannon1948,jaynes1957information,sivia2006data} guides us to logical forms for the prior distributions $p(N)$ and $p(\theta)$ assuming there's some known prior average target population size, $\hat{N}$, used for campaign planning, like the one from GRID3 \cite{grid3website} mentioned previously. For completeness in this section, we take
\begin{align}
    p(N) &= \frac{1}{\hat{N}}\left(1 - \frac{1}{\hat{N}}\right)^{N-n_1}\label{eq:N_prior}\\
    p(\theta) &= \frac{1}{2}\text{csch}^2\left(\frac{\theta}{2}\right)\left[\frac{\theta}{2}\text{coth}\left(\frac{\theta}{2}\right) -1 \right]\label{eq:t_prior},
\end{align}
on domains $N\geq n_1$ and $\theta \in (-\infty,\infty)$. To whatever extent it's helpful, Eq. \ref{eq:N_prior} might be called a left-truncated geometric distribution. Eq. \ref{eq:t_prior} doesn't have a name to our knowledge but is an exact consequence of $0 \leq \pi_0, \pi_1 \leq 1$ and the definition of the log odds ratio.

The joint prior distribution $p(N)p(\theta)$ is log-contour-plotted (black lines) in Fig. \ref{fig:inf1}'s first panel. The distribution is peaked at $N=n_1$, $\theta = 0$, and then falls geometrically with increasing $N$ and exponentially with $\theta$. Intuitively, the prior encodes that the target population is at least $n_1$ and otherwise finite. Meanwhile, the symmetry around $\theta=0$ represents our ignorance on how $R_i = 1$ is associated with campaign coverage in the absence of any data.

The observed $\Delta = 0.386 - 0.927$ in Table \ref{tab:data} specifies the likelihood $p(\Delta\,|\,N,\theta)$, which we visualize as a tan shaded interval within $5\%$ of the log-maximum in Fig. \ref{fig:inf1}. The likelihood defines a ridge of $N$, $\theta$ pairs that are consistent with $\Delta$. As $\theta$ increases to infinity, OD children are infinitely more likely than ZD children to participate in the campaign, and we see the ridge approach the perfect OD coverage estimate, $N\approx490$k, discussed above.

But with the machinery in Eq. \ref{eq:inf} we can solve the inference problem more completely, as illustrated in Fig. \ref{fig:inf1}'s second panel. Log joint posterior contours show how geometric features of the inputs combine to highlight the region most probably consistent with the data and prior uncertainties. The intuition that both perfect OD coverage and extremely high $N$ are unlikely leads us to the corner in the likelihood ridge. Marginal distributions $p(N\,|\,\Delta)$ and $p(\theta\,|\,\Delta)$, estimated via numerical integration and visualized along the corresponding axes, highlight the specific $\theta$ and $N$ values most consistent with the first two rows in Table \ref{tab:data}. 

Overall, in Kano in 2023, if we believe the tally sheets, we also must believe in a large, significantly less accessible ZD population that was missed by the campaign.

\section{Recapture coverage}

\begin{figure}[b]
\centering\includegraphics[width=\linewidth]{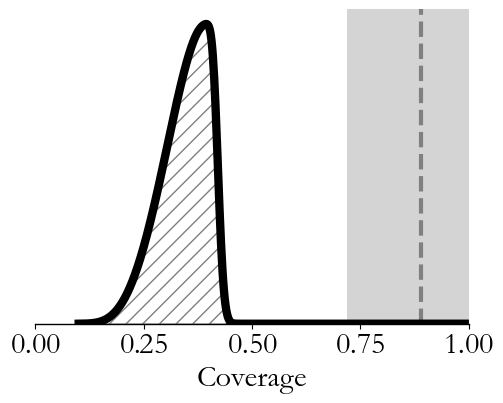}
\caption{\label{fig:cov1} Recapture coverage. Taking the tally sheets as they are, we get a campaign coverage estimate (black) that's inconsistent with the corresponding post-campaign survey (grey).}
\end{figure}

Getting from Fig. \ref{fig:inf1} to a coverage estimate is straight-forward since campaign coverage is $n_1/N$, a one-to-one function of $N$. In other words, we can take the marginal distribution $p(N\,|\,\Delta)$ and compute $p(n_1/N\,|\,\Delta)$ directly. The result (black) is visualized alongside the corresponding PCCS estimate (grey) in Fig. \ref{fig:cov1}.

Based on the tally sheets, our best guess for campaign coverage is $n_1/N^* \approx 39\%$, with a hard stop at the $42\%$, $\psi\rightarrow\infty$ limiting estimate. There is essentially no chance the $89\%$ PCCS coverage estimate ($95\%$ interval shaded) is consistent with the recapture approach. 

To add some texture to this result, we can calculate an individual-level contingency table $p(R_i,S^1_i\,|\,\Delta)$ encapsulating the implications consistent with the tally sheets. Specifically, if we take the marginal probabilities
\begin{align*}
\sum_{S_i^1} p(R_i=1,S^1_i\,|\,\Delta)&=p(R_i=1\,|\,\Delta)\approx C_{n_0}(\mathbf{R}),\\
\sum_{R_i} p(R_i,S^1_i=1\,|\,\Delta)&=p(S^1_i=1\,|\,\Delta)\approx \frac{n_1}{N^*},
\end{align*}
we find
\begin{align}
    p(R_i,S^1_i\,|\,\Delta) = \begin{BNiceMatrix}[first-row,first-col]
                    & R_i=0 & R_i=1 \\
                    S_i^1=0 & 0.557 & 0.049 \\
                    S_i^1=1 & 0.057 & 0.337
                    \end{BNiceMatrix}\label{eq:cont1}
\end{align}
making use of the most probable log odds ratio in Fig. \ref{fig:inf1} and noting that $\pi_{j} \equiv p(S_i^1=1\,|\,R_i=j)$ for $j=0,1$ in this new notation. 

At the most probable parameter values, an OD child is more than 9 times as likely to be reached by the campaign than a ZD child. Along those lines, we estimate that the campaign covers $87.4\%$ of the OD population but only $9.3\%$ of the ZD population. The probability of a child getting their first vaccine from the campaign is $p(R_i = 0\,|\,S_i^1=1) \approx 14.5\%$. Meanwhile, in this picture of the 6 LGAs, we find $N^*\approx518$k, more than $3.5$ times the GRID3 estimate. 

Eq. \ref{eq:cont1} paints a surprising picture, particularly in light of the campaign's stated $95\%$ coverage goal \cite{nigeria2023pccs}. We should and will be skeptical of it. But before that, the PCCS results in Table \ref{tab:data} are still lingering, and it's interesting that surveyed coverage among ZD was $11\%$, in reasonable agreement with the recapture estimate.

It's helpful to think of participation in the PCCS as another length $N$ binary vector, $\mathbf{S}^2 = (S_1^2, S_2^2, ..., S_N^2)$ with $S_i^2=1$ indicating the $i$th child was surveyed and $\sum_i S^2_i = n_2$. While the PCCS has well-defined survey design, and is unbiased in principle like the DHS, Fig. \ref{fig:cov1} at least suggests we should test for potential issues. With that motivation, we might have two competing hypotheses for its sampling distribution $p(\mathbf{S}^2\,|\,\mathbf{S}^1,\mathbf{R})$ in the post-campaign context. 

The more diplomatic hypothesis is that $p(\mathbf{S}^2\,|\,\mathbf{S}^1,\mathbf{R}) \approx p(\mathbf{S}^2\,|\,\mathbf{R})$. In words, PCCS is conditionally independent of the campaign, and sampling bias is explained by health care access. Surveyed campaign coverage estimates are spuriously inflated because the campaign shares a correlation with $R_i = 1$. In this case, as we did for $\mathbf{S}^1$, we would test against
\begin{align*}
    \ell_1 \equiv \sum_{i=1}^N S^2_iR_i &\sim \text{Binom}\left\{NC_N(\mathbf{R}),\tilde{\pi}_1\right\},\\
    \ell_0 \equiv \sum_{i=1}^N S^2_i(1-R_i) &\sim \text{Binom}\left\{N(1-C_N(\mathbf{R})),\tilde{\pi}_0\right\},
\end{align*}
such that $\ell_0 + \ell_1 = n_2$ and the sampling probabilities $\tilde{\pi}_0$ and $\tilde{\pi}_1$ reduce to one unknown as in Eq. \ref{eq:nch_dist}.

On the other hand, the more cynical $p(\mathbf{S}^2\,|\,\mathbf{S}^1,\mathbf{R}) \approx p(\mathbf{S}^2\,|\,\mathbf{S}^1)$ has campaign participation itself as the dominant bias, maybe because incentive structures pressure programs to have high coverage campaigns. In that case we might test
\begin{align*}
    m_1 \equiv \sum_{i=1}^N S^2_iS_i^1 &\sim \text{Binom}\left\{\sum_{i=1}^N S_i^1,\hat{\pi}_1\right\},\\
    m_0 \equiv \sum_{i=1}^N S^2_i(1-S_i^1) &\sim \text{Binom}\left\{N - \sum_{i=1}^N S_i^1,\hat{\pi}_0\right\},
\end{align*}
a comparable model, with $m_0+m_1=n_2$, but with the population structured by campaign participation instead of RI coverage, and so with distinct sampling probabilities, $\hat{\pi}_0$ and $\hat{\pi}_1$, in general.

In the first case, campaign coverage is 
\begin{align}
    \begin{split}
    \text{E}\left[\frac{1}{n_2}\sum S_i^2 S_i^1\right] &= \frac{\text{E}[\ell_1]}{n_2}p(S^1_i=1\,|\,R_i=1)\\
    &+\left(1-\frac{\text{E}[\ell_1]}{n_2}\right)p(S^1_i=1\,|\,R_i=0),
    \end{split}\label{eq:pccs_cov1}
\end{align}
a weighted average of OD and ZD coverage. Eq. \ref{eq:pccs_cov1} can be compared to the overall estimate in Table \ref{tab:data} to specify $\text{E}[\ell_1]/n_2$ consistent with Eq. \ref{eq:cont1}. Meanwhile, in the second case, surveyed campaign coverage is $\text{E}[m_1]/n_2$ directly. In other words, both models can explain the overall PCCS estimate with different underlying mechanisms.

But with each sampling distribution specified by the $89\%$ number in Table \ref{tab:data}, we can calculate the statistic $F = \sum_i S^2_iS^1_i(1-R_i)/n_2$, the surveyed estimate that a child gets their first measles vaccine from the campaign. In the $p(\mathbf{S}^2\,|\,\mathbf{R})$ model, this is just the second term in Eq. \ref{eq:pccs_cov1}, and we would estimate $F < 2\%$ almost certainly. Meanwhile, in the $p(\mathbf{S}^2\,|\,\mathbf{S}^1)$ model, we find $F = (\text{E}[m_1]/n_2)\times p(R_i=0\,|\,S_i^1=1) \approx 12.9\%$, and highly likely bounded between $9$ and $15\%$, an order of magnitude higher. 

Reflecting on the two mechanisms, if the campaign and PCCS share an underlying correlation with $R_i=1$, surveying a ZD child who was also vaccinated in the campaign is rare twice over. But if the PCCS is correlated with $S_i^1=1$ directly, surveying a ZD child is much less of a hindrance. In other words, if we believe the tally sheets from 2023, we must also believe that the corresponding PCCS was biased towards high coverage directly.

\section{Some asymptotic intuition}

So far we've discussed one explanation for the discordance in Table \ref{tab:data}, one that takes the DHS and the tally sheets as the two most reliable pieces of data. It's healthy to ask how potential recording errors or biases might influence our results. With that motivation in this section, we derive a coarse but interpretable approximation to $n_1/N^*$, which highlights how the recapture estimates change with distortions to the data.

In Appendix \ref{ap:A}, where we discuss some known statistical properties of the distribution in Eq. \ref{eq:nch_dist}, we show that its mode, $k_1^*$, is related to $\theta$ through
\begin{align}
    \frac{k_1^*}{n_1} = \frac{A(\theta)}{2n_1(e^\theta -1)}\left[1-\frac{B(\theta)}{A(\theta)}\right],\label{eq:mode}
\end{align}
where 
\begin{align*}
    A(\theta) &= N + (e^\theta-1)(n_1 + N C_N(\mathbf{R})),\\
    B(\theta)^2 &= A(\theta)^2 - 4e^\theta(e^\theta-1)n_1NC_N(\mathbf{R}).
\end{align*}
In our case, reflecting on Fig. \ref{fig:inf1}, we're in a situation where campaigns are much better at reaching OD children than ZD children. Mathematically then, we're concerned with inference stability in the large $\theta$ limit, which means characterizing Eq. \ref{eq:mode} to leading order in $e^{-\theta} = 1/\psi$ and calculating the downstream properties of Eq. \ref{eq:inf}. 

To start, we expand the terms outside and inside the brackets in Eq. \ref{eq:mode} separately, and find
\begin{align*}
    \frac{A(\theta)}{2n_1(e^\theta -1)} = \frac{1}{2}\left(1 + \frac{C_N(\mathbf{R})}{z} + \mathcal{O}\left(e^{-\theta}\right)\right),
\end{align*}
where we've introduced the notation $z \equiv n_1/N$, and
\begin{align*}
    \frac{B(\theta)}{A(\theta)} = \sqrt{1 - \frac{4zC_N(\mathbf{R})}{(z+C_N(\mathbf{R}))^2}} + \mathcal{O}\left(e^{-\theta}\right).
\end{align*}
Combining the two expressions and collecting terms exposes the mode's leading order behavior,
\begin{align*}
    \frac{k_1^*}{n_1} = \frac{C_N(\mathbf{R})}{z} + \mathcal{O}\left(e^{-\theta}\right),
\end{align*}
recovering the $\psi\rightarrow \infty$ result discussed in the last two sections, that $k_1^* \rightarrow NC_N(\mathbf{R})$, the full OD population. 

When $n_1$ is large, Eq. \ref{eq:nch_dist} is sharp, and the mean is close enough to the mode for our purposes here. Taking $\text{E}[k_1]/n_1 \approx C_N(\mathbf{R})/z$, Eq. \ref{eq:inf} reduces to
\begin{align}
    \begin{split}
    \ln p(z,\theta\,|\,\Delta) \approx& \,\,\text{const.} + \frac{n_1}{z}\ln\left(1-\frac{1}{\hat{N}}\right)\\
    &-\frac{1}{2\text{V}[\Delta]}\left(\Delta - C_N(\mathbf{R})\left[1 - \frac{1}{z}\right]\right)^2,\label{eq:logpost}
    \end{split}
\end{align}
where the constant depends on $\theta$ but not on $z$. Then, in the same large $n_1$ limit, $\text{V}[\Delta]\approx\text{V}[C_{n_0}(\mathbf{R})]$, and Eq. \ref{eq:logpost} can be maximized to find $z^*$, the most probable campaign coverage estimate. Carrying out the calculation, with $\ln(1-1/\hat{N})\approx-1/\hat{N}$, we find
\begin{align}
    \frac{1}{z^*} = \frac{N^*}{n_1} \approx 1 - \frac{\Delta}{C_N(\mathbf{R})}-\frac{n_1}{\hat{N}}\frac{\text{V}[C_{n_0}(\mathbf{R})]}{C_N(\mathbf{R})^2}.\label{eq:approx_cov}
\end{align}

Eq. \ref{eq:approx_cov} bounds coverage in the $\psi\rightarrow\infty$ limit and helps scaffold our thinking. The ratio $\text{V}[C_{n_0}(\mathbf{R})]/C_{N}(\mathbf{R})^2$ is an index of dispersion, quantifying our uncertainty in RI coverage. It decays as $1/n_0$ for the DHS. As a result, when our RI coverage estimate is precise and the prior population estimate is $\mathcal{O}(n_1)$, the final term in Eq. \ref{eq:approx_cov} is small. In 2023 in Kano for example, it's $-0.006$ if the DHS is genuinely unbiased so $C_{n_0}(\mathbf{R}) \approx C_{N}(\mathbf{R})$.

If the DHS were biased, a possible error neglected in the previous two sections, we might more generally write $C_{n_0}(\mathbf{R}) = (1+\alpha)C_N(\mathbf{R})$ with $\alpha\neq0$ characterizing the degree. In that case, a precise but shifted DHS estimate would modify $z^*$ like so,
\begin{align*}
    z^* \approx \frac{C_{n_0}(\mathbf{R})}{C_{n_1}(\mathbf{R})}\left[1-\alpha\left(1-\frac{C_{n_0}(\mathbf{R})}{C_{n_1}(\mathbf{R})}\right)\right],
\end{align*}
to leading order in $\alpha$. Realistically, if it's biased at all, the DHS likely overestimates $C_N(\mathbf{R})$, so $\alpha>0$. In the large $\theta$ regime we're in, that would decrease $z^*$, making the discrepancy in Fig. \ref{fig:cov1} worse. DHS bias doesn't seem like a viable explanation for the issues in Table \ref{tab:data}.

Meanwhile, when it comes to tally sheet recording, Eq. \ref{eq:approx_cov} tells us that the recapture coverage estimate depends mainly on the ratio of RI coverage to the tally sheet average. In the Kano example, an error that inflates $n_1$ by 10 times would modify recapture coverage by only $0.1\%$ if it maintains $C_{n_0}(\mathbf{R})/C_{n_1}(\mathbf{R})$. That said, those same types of errors effect $N^*$ directly, and we should interpret recapture target population estimates with that in mind. 

\section{A hypothetical tally}

Eq. \ref{eq:approx_cov} suggests that the recapture coverage estimate would align with Kano's PCCS if $C_{n_1}(\mathbf{R})$ was roughly halved. In fact, it seems like the recapture estimate is robust in the sense that no other change will do the job. 

Tabling discussions of plausibility for now, let's assume half of the OD tallies were misrecorded and were actually ZD. In that case $C_{n_1}(\mathbf{R})$ goes from $92.7\%$ to $46.4\%$, and we can calculate the consequences from Eq. \ref{eq:inf} onward. 

\begin{figure}
\centering\includegraphics[width=\linewidth]{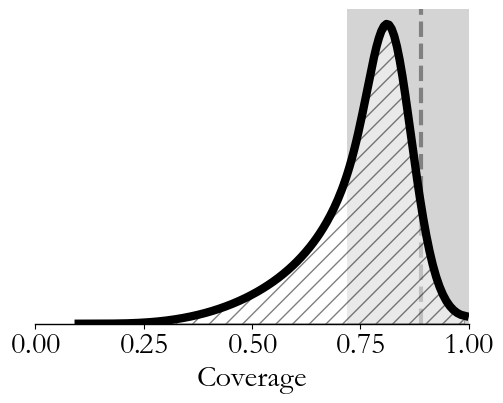}
\caption{\label{fig:cov2} Hypothetical tally sheets in Kano. Recapture coverage (black) and the post-campaign survey (grey) agree if half of the OD tallies are actually ZD.}
\end{figure}

The new coverage estimate is visualized in Fig. \ref{fig:cov2}, and it agrees with the PCCS as expected. Revisiting the contingency table, $p(R_i, S_i^1\,|\,\Delta)$, at the most probable estimates, we find
\begin{align}
    p(R_i,S^1_i\,|\,\Delta) = \begin{BNiceMatrix}[first-row,first-col]
                    & R_i=0 & R_i=1 \\
                    S_i^1=0 & 0.139 & 0.051 \\
                    S_i^1=1 & 0.475 & 0.335
                    \end{BNiceMatrix}\label{eq:cont2}.
\end{align}
Now, an OD child is 1.1 times as likely as a ZD child to be reached by the campaign, the campaign covers $86.8\%$ of the OD population and $77.4\%$ of the ZD population, and the probability of a child getting their first vaccine from the campaign is $59\%$. The estimate of target population size changes as well, $N^*\approx252$k, a potentially more reasonable \cite{lang2025global} $1.7$ times the GRID3 estimate. 

Our previous thoughts on PCCS sampling basically reverse. Specifically, to explain overall coverage surveyed around $89\%$, many more ZD children are required to have been vaccinated in the campaign. Under the $p(\mathbf{S}^2\,|\,\mathbf{S}^1)$ sampling model specified by PCCS coverage, we'd estimate that $F > 37\%$ of children surveyed got their first measles vaccine from the campaign. Meanwhile, in the competing $p(\mathbf{S}^2\,|\,\mathbf{R})$ model, $F$ is unconstrained, covering the $11\%$ number in Table \ref{tab:data}. So in this hypothetical situation, the campaign and the PCCS sharing an underlying correlation with $R_i=1$ seems necessary. 

\section{Some structured error models}

Fig. \ref{fig:cov2} demonstrates that it's possible for the recapture and PCCS estimates to agree. Realizing that agreement in Kano would require a systematic recording error where about half of OD tallies are actually ZD. It seems unlikely that a net $50\%$ false positive rate is realistic, but it's worth ruling out more thoroughly.

One potential mechanism is caregiver recall -- maybe children get other vaccines and their parents get them mixed up when asked about measles. Literature on this topic \cite{MODI20184161} doesn't suggest a large, one directional effect however. In most studies, people are roughly as likely to forget doses they received, falsely reporting ZD, as they are to invent doses they haven't. 

Recall doesn't seem affected by the campaign environment either. Looking at serological data from the 2020 measles campaign in Zambia \cite{10.1371/journal.pgph.0003209}, differences in seropositivity between those with documented vaccine and those with recalled vaccine suggests that OD memory is accurate $\approx86\%$ of time. That number is in good agreement with more focused recall studies \cite{gareaballah1989accuracy}.

Moving on then to recording itself, we might guess that some vaccination teams didn't bother asking about dose history. Across the 409 teams in the 6 LGAs, 100 never tally a ZD in any age group. If we remove those tally sheets from the dataset however, $C_{n_1}(\mathbf{R}) \rightarrow 0.906$. Not even close to what we'd need.

We can try more sophisticated misrecording indicators too. Since the number of children who show up on a given day is a random process, we shouldn't expect any preference for round numbers. Looking at the work across the teams, with 10 numbers per sheet, we get a sample of up to 90 nonzero numbers per team across the campaign. We can test for example if the last digit in each nonzero entry is uniformly drawn from 0 to 9 using a standard Pearson's $\chi^2$ test applied to each team's tallies.

There's definitely evidence of non-uniform recording. Some examples are striking: One team recorded 33 non-zero numbers across the campaign, 30 of which ended in 0. More broadly, if we remove the sheets from teams with less than $10\%$ probability of drawing from a uniform last digit distribution, we lose the work from $60\%$ of the vaccinators. Doses delivered in the campaign decline substantially, $n_1 \rightarrow 77139$, but the OD fraction remains essentially the same, $C_{n_1}(\mathbf{R}) \rightarrow 0.923$. As a result, as expected from Eq. \ref{eq:approx_cov}, the recapture coverage estimate is  unchanged, but $N^*\rightarrow229$k, larger than the GRID3 estimate by 1.6 times.

Returning to the motivating goal, while we could probably find more reasons to question subsets of the tally sheets, it's difficult to imagine a mechanism for systematic false OD tallies at such a large scale. From an estimate stability perspective then, the recapture coverage approach seems generally reliable.

\section{Conclusion}

At this stage, reflecting on Table \ref{tab:data}, the tally sheets give compelling evidence that Kano's 2023 campaign had lower coverage in the 6 LGAs than the PCCS's state-wide $89\%$. At the same time, the tally sheets imply there are more than 3 times as many 1 year-olds as people think, so it also seems likely that the number of tallies is inflated in some way. To what extent is unclear. 

More to the point though, we have shown that implementation data, despite being collected outside a traditional survey platform, can power provably robust inferences about campaign quality, target population size, and measurement. The approach illustrated in Fig. \ref{fig:inf1} is applicable instantly after tallies are totaled, Eq. \ref{eq:approx_cov} suggests that resolution is limited mainly by precision in RI coverage, and results like Eq. \ref{eq:cont1} meaningfully highlight campaign failure modes. Analysis like this costs very little. It should probably be a part of all campaigns.

If it were, it would inspire more thoughtful tally sheet design and more intentional data collection. We might reconsider PCCS's role in that context. The Kano example suggests that PCCS is less suited to measuring coverage, and is better for measuring other campaign details, like the characteristics of children conditional on their campaign participation. It seems worth some study to design post campaign surveys with that narrower focus in mind, giving vaccination programs more fit-for-purpose measurement tools.

\begin{acknowledgments}
This work benefited immensely from discussions with colleagues at the Gates Foundation. In particular, I want to thank Amine El Mourid, Kevin McCarthy, Roy Burstein, Hil Lyons, Katie Maloney, Chris Wolff, Marisa Gaetz, and Ana Leticia Nery. Their technical and programmatic perspectives influenced this research at every stage. I also want to thank GRID3 for their care with the pilot study, and their input on early iterations of this work. Finally, I want to thank the National Primary Health Care Development Agency in Nigeria for their role in pilot study data collection.
\end{acknowledgments}

\appendix
\section{Noncentral, hypergeometric likelihoods}\label{ap:A}

Both the campaign and PCCS are modeled as two binomial samples constrained in sum, as in Eq. \ref{eq:nch_dist}. That construction is called Fisher's noncentral hypergeometric distribution, and its properties are well-studied, but the derivations in the literature are a little confusing with results are scattered across multiple books and papers \cite{McCullaghNelder1989,Cornfield1956,Levin1984,EisingaPelzer2011}. We include a concise derivation of the asymptotic results used here in case it's useful for reference.

A random variable $x$ is a noncentral hypergeometric if
\begin{align*}
    p(x\,|\,m,N_1,N_2,\psi) \propto \binom{N_1}{x}\binom{N_2}{m-x}\psi^x,
\end{align*}
where $\psi$ is the odds ratio of contributions to the $m$ samples coming from subpopulation $N_1$ relative to $N_2$. Taking the log, applying Sterling's approximation to the binomial coefficients, and maximizing gives the condition
\begin{align}
    \psi = \frac{x^*(N_2 - m + x^*)}{(N_1 - x^*)(m-x^*)}\label{eq:mode_cond}
\end{align}
which is quadratic in the mode $x^*$ and solvable as in Eq. \ref{eq:mode}. Differentiating the log-probability with respect to $x$ again and evaluating at $x^*$ gives
\begin{align*}
    \tilde{\text{V}}[x] = \left[\frac{1}{x^*} + \frac{1}{N_1 - x^*}+ \frac{1}{m - x^*} + + \frac{1}{N_2 - m + x^*}\right]^{-1}
\end{align*}
for the variance around the mode. Frequently in asymptotic Bayesian inference, we stop here \cite{sivia2006data}, but for this distribution we can make these approximations more accurate with two more ideas.

The first is credited to McCullagh and Nelder \cite{McCullaghNelder1989}, who noted that $p(x\,|\,m,N_1,N_2,\psi)$ should reduce to the standard (central) hypergeometric distribution when $\psi=1$. Comparing the mode and variance above to that distribution's known statistics inspires a correction
\begin{align*}
    \text{V}[x] = \frac{N_1+N_2}{N_1+N_2-1}\tilde{\text{V}}[x],
\end{align*}
which they demonstrate improves accuracy.

The second improvement is to adjust the mode closer to the mean \cite{Levin1984}, which is what's actually needed in Eq. \ref{eq:inf}. The mode condition Eq. \ref{eq:mode_cond} inspires us to calculate the averages $\text{E}[x(N_2-m+x)]$ and $\text{E}[(N_1-x)(m-x)]$. Manipulations of the binomial coefficients expose the exact formula,
\begin{align*}
    \psi = \frac{\text{E}[x(N_2-m+x)]}{\text{E}[(N_1-x)(m-x)]},
\end{align*}
generally credited to Cornfield \cite{Cornfield1956}. Rearranging gives the mean-variance relationship,
\begin{align*}
    f(\text{E}[x]) + (\psi-1)\sigma^2 = 0,
\end{align*}
where $\sigma^2 = \text{E}[x^2] - \text{E}[x]^2$ is the exact variance around the mean, and we've defined
\begin{align*}
    f(\xi) \equiv \psi(N_1-\xi)(m-\xi) - \xi(N_2 - m +\xi).
\end{align*}
Note that Eq. \ref{eq:mode_cond} implies $f(x^*) = 0$, so we can take $f(\text{E}[x]) \approx f'(x^*)(\text{E}[x]-x^*)$ to perturb the mode towards the mean. Carrying out the calculation gives
\begin{equation*}
    \text{E}[x] \approx x^* + \frac{(\psi-1)\sigma^2}{(\psi-1)[N_1+m-2x^*] + N_1 + N_2},
\end{equation*}
which we evaluate in practice with $\sigma^2 \approx \text{V}[x]$ from above. These are the two statistics of Eq. \ref{eq:nch_dist} needed in the Gaussian approximation to the likelihood $p(\Delta\,|\,N,\theta)$ used throughout the main text. 

It's worth briefly pointing out that we approximate $p(C_{n_0}(\mathbf{R})\,|\,n_0)$ as Gaussian too. That approximation makes errors at the level of tail probabilities where $C_{n_0}(\mathbf{R}) < 0$ or $C_{n_0}(\mathbf{R}) > 1$ have nonphysical support. The total tail probability is 
\begin{align*}
    \mathcal{E} = 1 - \int_0^1dz\,\mathcal{N}(z\,|\,\text{E}[C_{n_0}(\mathbf{R})],\text{V}[C_{n_0}(\mathbf{R})]),
\end{align*}
and for the DHS in Kano, numerical integration gives $\mathcal{E}\approx0$, as expected for a well-powered survey. In other words, we can safely use the Gaussian likelihood in our example. But notably, if we were working at higher spatial resolution with correspondingly larger $\text{V}[C_{n_0}(\mathbf{R})]$, $\mathcal{E}$ could potentially be significant.

\section{Maximum entropy prior probabilities}\label{ap:B}

The inference problem in Eq. \ref{eq:inf} requires prior structure through $p(N,\theta) = p(N)p(\theta)$. Fig. \ref{fig:inf1} makes it clear that this choice is influential since the likelihood alone, the tan ridge, does not have a distinct maximum. Different choices for $p(N,\theta)$ will yield different inferences, a reflection of those choices expressing higher preference for some $N$ or $\theta$ than that in Eqs. \ref{eq:N_prior} and \ref{eq:t_prior}. 

As mentioned in the main text, we calculate information entropy maximizing priors consistent with the data available in advance of the campaign. This is a theoretically supported and rigorous approach to choosing $p(N,\theta)$, especially in problems where prior information is a key ingredient \cite{Shannon1948,jaynes1957information}. Maximum entropy is a principle that helps us find distributions consistent with the data we have but with the least amount of structure otherwise. Ref. \onlinecite{sivia2006data} is an approachable introduction.

In any case, starting with $\theta$, we first consider a sampling probability $\pi_j$ with $j=0,1$ as in Eq. \ref{eq:nch_dist}. If we think of a $p(\pi_j)$, the related entropy is
\begin{align*}
    H_{\pi}[p(\pi_j)] = -\int_0^1 d\pi_j \,p(\pi_j)\ln(p(\pi_j)),
\end{align*}
reflecting the fact that $0\leq \pi_j \leq 1$. We want to maximize $H_{\pi}$ subject to the constraint that $p(\pi_j)$ is normalized. 

Jaynes proved \cite{jaynes1957information}, by discretizing the 0 to 1 interval into $m$ pieces, that $p(\pi_j) \approx 1/m$, which yields $\pi_j \sim \text{Uniform}\{0,1\}$ exactly in the continuous limit. This is the intuitive result. Knowing nothing else, just that $\pi_j$ is a probability, we'd place it anywhere from 0 to 1 without preference. 

To specify $p(\theta)$, we need to work from $\pi_0$ and $\pi_1$ to a log odds ratio. The odds, $y_j = \pi_j/(1-\pi_j)$, is an invertible transformation, so
\begin{align*}
    p(y_j) &= \left|\frac{d\pi_j}{dy_j}\right|\,p\left(\pi_j = \frac{y_j}{1+y_j}\right)\\
    &= \frac{1}{(1+y_j)^2},
\end{align*}
which is apparently called a beta-prime distribution. Moving to $z_j = \ln y_j$ is another invertible transformation, which this time yields
\begin{align*}
    p(z_j) = \frac{1}{(1+e^{z_j})(1+e^{-z_j})},
\end{align*}
which is a product of logistic functions evaluated at $\pm z_j$. The density $p(z_j)$ is sometimes called a generalized logistic distribution in light of that fact. 

In this notation, $\theta = \ln z_1 - \ln z_0$, and so $p(\theta)$ is a convolution of two generalized logistic distributions. To evaluate the convolution, we first calculate the characteristic function,
\begin{align*}
    \varphi_z(t) \equiv \text{E}\left[e^{itz}\right] = \int_{-\infty}^{\infty} dz \frac{e^{itz}}{(1+e^{z})(1+e^{-z})},
\end{align*}
where $i = \sqrt{-1}$ and we've suppressed the index on $z$. The integral can be taken on a semi-circular contour in the complex $z$ plane, summing over the poles on the imaginary axis. We find $\varphi_z(t) = \pi t \text{csch}(\pi t)$, implying
\begin{align*}
    \varphi_{\theta}(t) &= \varphi_{z_1}(t)\varphi_{z_0}(-t)\\
    &= (\pi t)^2 \text{csch}^2(\pi t),
\end{align*}
via the convolution theorem. Taking the inverse Fourier transform, again by summing residues within a semicircular contour, gives
\begin{equation*}
    p(\theta) = \frac{1}{2}\text{csch}^2\left(\frac{\theta}{2}\right)\left[\frac{\theta}{2}\text{coth}\left(\frac{\theta}{2}\right) -1 \right],
\end{equation*}
which is Eq. \ref{eq:t_prior} in the main text. Note that $p(\theta)$ is a direct consequence of the definition $e^\theta = \pi_1(1-\pi_0)/(1-\pi_1)\pi_0$, and so $p(\theta)$ maximizes entropy exactly.

\begin{figure}
    \centering
    \includegraphics[width=\linewidth]{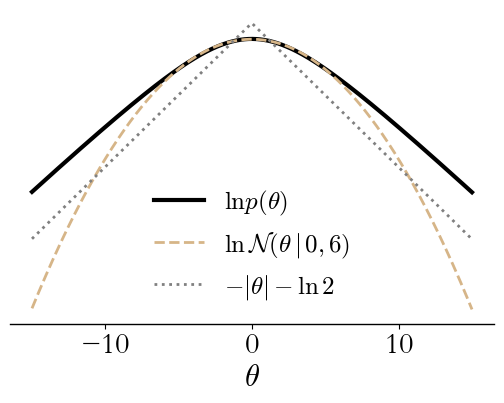}
    \caption{Interpreting Eq. \ref{eq:t_prior}. Extreme values of log odds ratios, $\theta$, aren't meaningfully different from one another.}
    \label{fig:t_prior}
\end{figure}

The result above is admittedly a little opaque. To build some intuition, we can expand $\ln p(\theta)$ to quadratic order around $\theta=0$. That teaches us that $p(\theta) \approx \mathcal{N}(\theta\,|\,0,6)$, a centered normal distribution with variance $6$. The comparison is visualized in Fig. \ref{fig:t_prior}, showing that the approximation is good for most of the probable domain, with heavier but still exponential tails in $p(\theta)$. Evidently odds ratios start to lose meaning above or below $\approx e^{\pm5}$, as one or both of the $\pi_i$ get closer and closer to 0 or 1. 

Moving on to $N$, we assume campaign planning is based on some prior estimate $\hat{N}$, which we model as the expected value of $p(\hat{N})$. Thus, we want to maximize 
\begin{align*}
    H_N[p(N)] = -\sum_{N=1}^\infty p(N)\ln p(N),
\end{align*}
subject to constraints 
\begin{align*}
    \sum_{N=1}^\infty p(N) &= 1\\
    \sum_{N=1}^\infty N p(N) &= \hat{N}.
\end{align*}
The usual Lagrange multiplier method \cite{jaynes1957information} gives
\begin{align*}
    p(N) = \frac{1}{\hat{N}}\left(1 - \frac{1}{\hat{N}}\right)^{N-1},
\end{align*}
a standard geometric distribution. At the moment, this result is unconditional on $n_1$, the total number of doses given in the campaign. In the context of Eq. \ref{eq:inf}, we need to incorporate that $N \geq n_1$. Truncating $p(N)$ by taking
\begin{align*}
    p(N\,|\,n_1) &= \frac{p(N)}{1 - \sum_{\tilde{N}=1}^{n_1-1}p(\tilde{N})},
\end{align*}
gives
\begin{align*}
    p(N\,|\,n_1) = \frac{1}{\hat{N}}\left(1 - \frac{1}{\hat{N}}\right)^{N-n_1},
\end{align*}
which is Eq. \ref{eq:N_prior} in the main text.

We note briefly that in principle GRID3 and others who produce $\hat{N}$ also occasionally publish measures of uncertainty. That said, the literature suggests that uncertainty in population sizes is poorly estimated \cite{lang2025global}, particularly at higher resolution. Moreover, that uncertainty does not inform campaign planning in the same way prior targets do. Regardless, we choose to ignore prior population uncertainty estimates to emphasize the values consistent with the tally sheets. But if we had stronger feelings on $\hat{N}$, they would manifest in $p(N\,|\,n_1)$ having faster decaying tails, which would effect the results. 

\bibliography{references}

\end{document}